\pdfoutput=1

\documentclass[11pt]{article}

\usepackage[preprint]{acl}

\usepackage{times}
\usepackage{latexsym}
\usepackage{amsmath}
\usepackage{graphicx}
\usepackage[T1]{fontenc}
\usepackage{amssymb}
\usepackage{booktabs}
\usepackage{enumitem}
\usepackage{multirow}
\usepackage{bbding}
\usepackage{pifont}
\usepackage{makecell}

\usepackage[utf8]{inputenc}

\usepackage{microtype}

\usepackage{inconsolata}

\title{Text-to-Song: Towards Controllable Music Generation Incorporating Vocals and Accompaniment}

\author{Zhiqing Hong, Rongjie Huang, Xize Cheng, Yongqi Wang, Ruiqi Li, Fuming You,\\ {\bf Zhou Zhao\dag, Zhimeng Zhang}\\
         Zhejiang University\\ \{zhiqinghong, rongjiehuang, zhaozhou\}@zju.edu.cn}

\begin{document}
\maketitle

\begin{abstract}
A song is a combination of singing voice and accompaniment. However, existing works focus on singing voice synthesis and music generation independently. Little attention was paid to explore song synthesis. In this work, we propose a novel task called text-to-song synthesis which incorporating both vocals and accompaniments generation. We develop Melodist, a two-stage text-to-song method that consists of singing voice synthesis (SVS) and vocal-to-accompaniment (V2A) synthesis. Melodist leverages tri-tower contrastive pretraining to learn more effective text representation for controllable V2A synthesis. A Chinese song dataset mined from a music website is built up to alleviate data scarcity for our research. The evaluation results on our dataset demonstrate that Melodist can synthesize songs with comparable quality and style consistency. Audio samples can be found in \url{https://text2songMelodist.github.io/Sample/}.
\end{abstract}
\section{Introduction}

Songs, as intricate musical compositions, have always enjoyed the greatest popularity among music enthusiasts. It inspires the pursuit of song synthesis by leveraging machine learning and artificial intelligence algorithms. It makes sense to generate a song conditioned on text modality (music score, natural language prompt, etc.). However, there is little exploratory research on text-to-song synthesis to our knowledge.

There are two related tasks. The first is singing voice synthesis, which converts the music score (lyrics, notes, and duration) to the singing voice. Existing SVS models have achieved remarkable achievement regarding quality \cite{opensinger, diffsinger,hong2023unisinger, zhang2022m4singer} and zero-shot ability \cite{qian2019autovc, casanova2022yourtts} but they can only generate vocals. Another similar task is music accompaniment generation \cite{ren2020popmag, dong2018musegan}, which usually aims at generating multi-track sequences in the symbolic domain or directly generating music waveform from text descriptions \cite{lyric2accom, yu2021conditional}. As presented in Figure \ref{fig:intro}, there are similarities among these three tasks, while notable distinctions exist. The accompaniments are often removed in data preprocessing to train an SVS model. And existing music generation models do not take vocals into account as the condition. Further exploration of text-to-song is inhibited. 

Neither serves as the suitable prior. To address this limitation, we propose a novel generative task, \emph{Text to Song}, which converts the music score (lyrics, notes, and duration) to the song, that is, singing voice with accompaniment. However, a text-to-song model is facing several challenges:

\begin{figure}[t]
    \centering
    \includegraphics[width=\columnwidth]{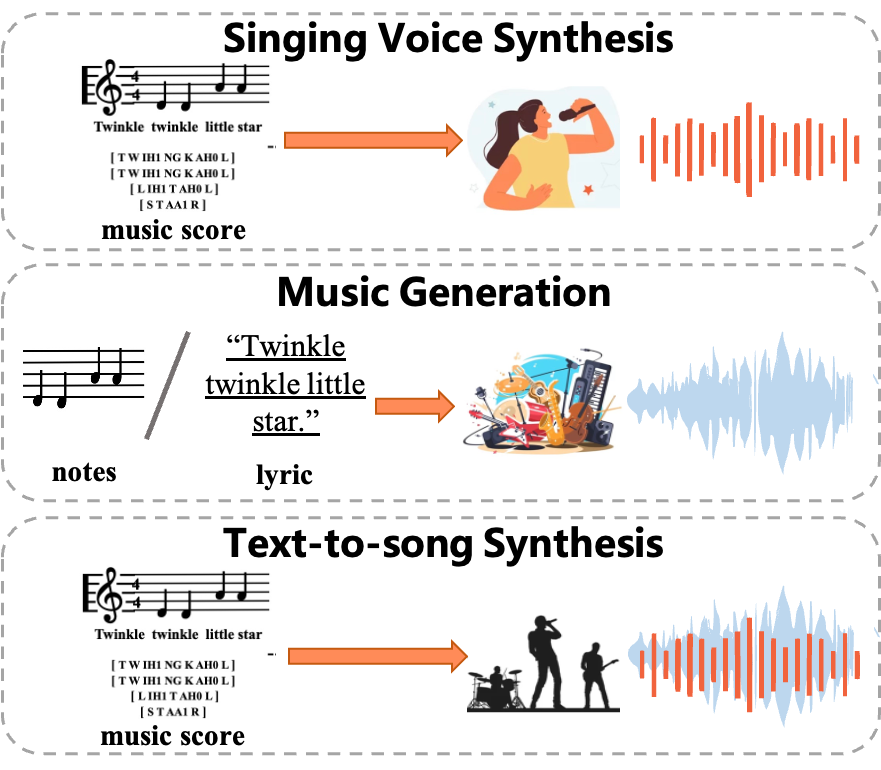}
    \caption{The comparison of three tasks: singing voice synthesis, accompaniment generation and text-to-song. In this work, We investigate on the relationship between vocal and accompaniment for text-to-song synthesis.}
    \label{fig:intro}
\end{figure}

\textbf{1) Process of Synthesis.} It is hard to achieve end-to-end generation since the song contains much more information (pitch variation, timbre, emotion, instruments, etc.) than the music score, which imposes a large burden on the model.

\textbf{2) Additional Control.} This is far from enough to model the diverse output while only feeding the music score to the text-to-song synthesis model. Some natural language prompts should be included as the condition to guide and control the accompaniment generation.

\textbf{3) Data Scarcity}. To the best of our knowledge, there is no dataset with pairs of vocal and accompaniment audios along with finely annotated music score (which should at least have lyrics transcription). It is the most intractable factor hindering research in this area.

In this paper, we propose Melodist, the first text-to-song model to generate music incorporating vocals and accompaniments from music score. To overcome the challenges mentioned above, we adopt several techniques: 1) Based on the human perception that the accompaniment complements the vocal melody, providing harmonic and rhythmic structure to enhance the overall musical expression, we introduce a two-stage text-to-song synthesis. Specifically, Melodist generates singing voice from the music score in Stage 1, then generates accompaniment given vocal in Stage 2. Finally, we mix the outputs of two stages to obtain the song. It releases the burden of our model to a large extent; 2) We utilize the attribute tags (mood, instruments, style, etc.) of each song segment and construct natural language prompts to guide the synthesis of the accompaniment. We further apply the Tri-Tower Contrastive Learning framework to extract better text representations; 3) We crawled some songs and the corresponding lyrics and tags related to attributes from music websites. We evaluate our model under different settings and the results demonstrate that Melodist can synthesize songs with comparable quality under the control of natural language prompts. 

The main contributions of our work can be summarized as follows:
\begin{itemize}
[left=0em,label=\textbullet]
    \setlength{\itemindent}{0pt}
    \setlength{\tabcolsep}{0pt}
    \setlength{\parskip}{0pt}
    \setlength{\partopsep}{0pt}
    \setlength{\itemsep}{0pt}
    \setlength{\topsep}{0pt}
    \setlength{\parsep}{0pt}
    \item We introduce a new task of text-to-song synthesis, which aims to convert the music score to the song incorporating vocal and accompaniment synthesis. We further propose Melodist, the first text-to-song model following two-stage song synthesis;
    \item We adopt natural language prompts to generate various types of accompaniment; 
    \item We design a tri-tower contrastive learning framework to connect the text context with its corresponding vocal and accompaniment pattern;
    \item We construct a dataset that provides not only pairs of vocals and accompaniment but also transcriptions in text format including lyrics and attribute tags. 
    \item We conduct extensive experiments to verify the effectiveness of Melodist. Experiment results show that Melodist exhibits high quality and great adherence.
\end{itemize}

\section{Related Work}
\subsection{Singing Voice Synthesis}
Substantial progress has been made in Singing Voice Synthesis (SVS). Several works \cite{singgan, kong2020hifi} have adopted generative adversarial networks (GANs) \cite{gan}, while there appear many end-to-end SVS models \cite{visinger, hong2023unisinger} based on variational autoencoder (VAE). DiffSinger \cite{diffsinger} is built on diffusion probabilistic models which can generate more high-fidelity outputs. In the realm of the Large Language Model recently, there are many emerging methods \cite{yang2023uniaudio, make-a-voice} modeling voice with an auto-regressive transformer in a compact and discrete space. However, these works discarded the accompaniments in data pre-processing, while we take accompaniment generation into account and investigate the relationship between vocals and accompaniments. 
\begin{figure*}[ht]
    \centering
    \includegraphics[width=\textwidth]{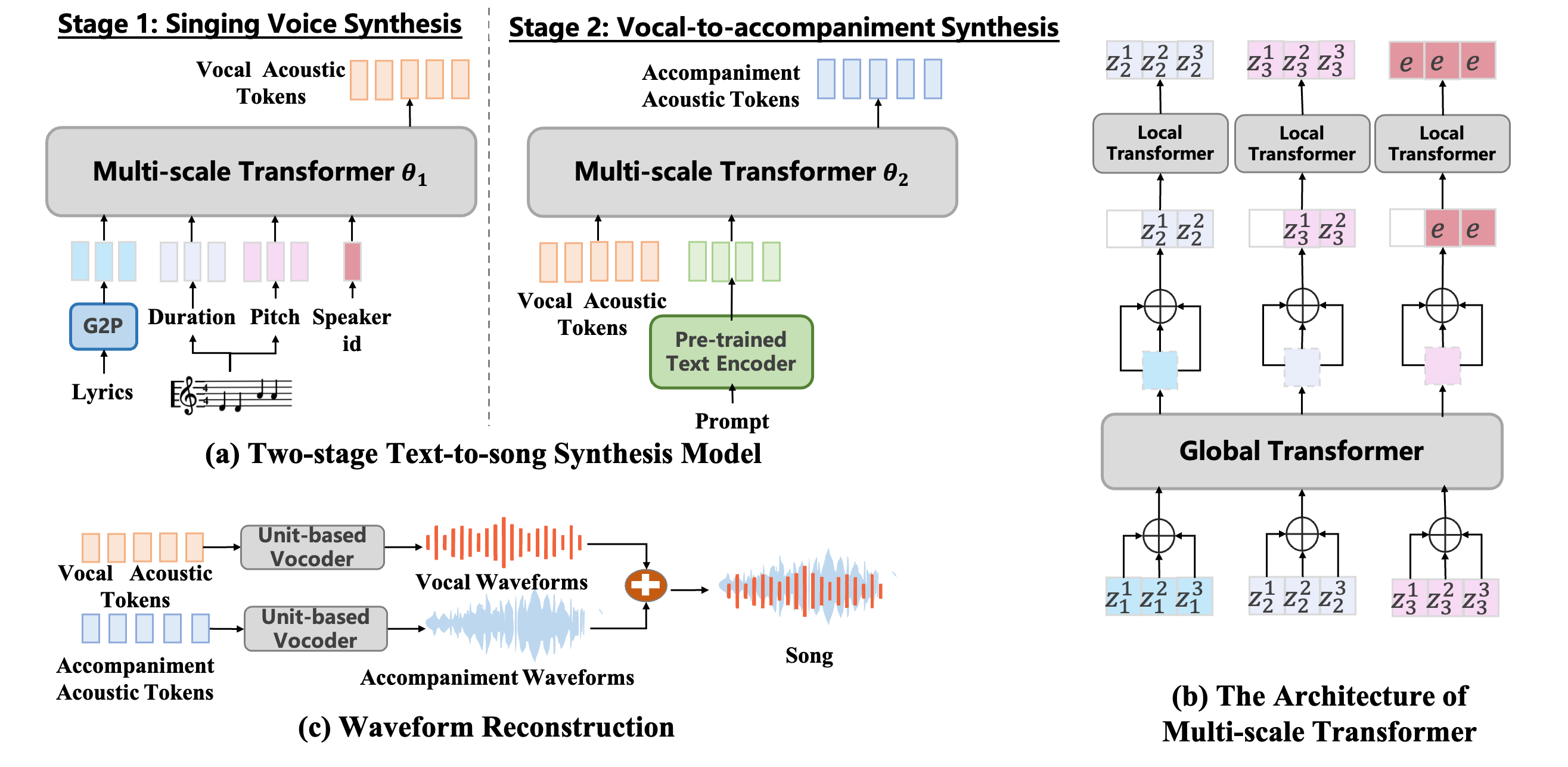}
    \caption{The overview of Melodist, the proposed two-stage text-to-song synthesis model. We present the two-stage pipeline in subfigure (a). In subfigure (b), we present the multi-scale Transformer architecture, in which e and $z_t^k$ denote <EOS> token and the k-th audio token at t-th frame, respectively.}
    \label{fig:overview}
\end{figure*}
\subsection{Accompaniment Generation}
Researchers on accompaniment usually work on musical symbolic tokens in a seq2seq setting. MuseGAN \cite{dong2018musegan} is the first model that generates multi-track polyphonic music with harmonic and rhythmic. There exist several works \cite{copet2023musicgen, agostinelli2023musiclm}trying to generate melody conditions on chord information for better music structure. Yang et al. \cite{yang2017midinet} designed MidiNet to generate melodies one bar after another. PopMAG \cite{ren2020popmag}was proposed to simultaneously generate five instrumental tracks in a single sequence. However, these methods rely highly on symbolic music representation. Recently, Donahue et al. presented SingSong \cite{donahue2023singsong}, a system that generates instrumental music to accompany input vocals. But the limitation remains in the lack of controllability related to mood, instruments, style, etc. In this work, we focus on developing a synthesis system that accepts natural language prompts for guiding the generation.

\subsection{Cross-modal Contrastive Learning}
Contrastive learning, which is first applied in computer vision domain \cite{clip, oord2018representation}, achieves high performance in many downstream tasks such as zero-shot recognition, image-text retrieval, etc. Along the same line in the audio domain, Wav2clip \cite{wu2022wav2clip} and Audioclip \cite{guzhov2022audioclip} are both derived from CLIP. To achieve more flexibility and generalization, CLAP \cite{elizalde2023clap,huang2023make} is proposed to learn audio concepts from natural language supervision instead of class labels. Recently, an increasing number of works \cite{chen2022learning, manco2022contrastive} exploring contrastive pre-training in the music domain. MuLan \cite{huang2022mulan} is the first model learning a joint embedding space for music and natural language trained with an unprecedented scale of weakly paired text and audio. In this work, we also leverage a contrastive learning framework to extract better text representations.

\section{Two-stage Text-to-song Synthesis}
In this section, we first present a formal Definition of text-to-song synthesis task. Then we will give an overview of the proposed model Melodist. Finally, we will elaborate on the approaches we adopt for controllable two-stage text-to-song synthesis. 

\subsection{Task Definition}
In this work, we present a novel task \emph{text-to-song} and extend it to controllable synthesis. Given the training set $D$ consists of $n$ data points $(s_i, p_i, c_i)$, $i=1,...,n$, where each element denotes a song, the description of its accompaniment and music score of its vocal, we convert the music score to song conditioned on the natural language prompt, which can be formulated as a conditional probability distribution modeling problem:

\begin{equation}
\begin{aligned}
p(S|C,P)=\prod \limits_{t=0}^T{p(s_t | s_{<t}, C, P; \theta)}
\end{aligned}
\end{equation}

Given that $S = S_v+S_a$, where $S$, $S_v$, $S_a$ denote song waveforms, vocal waveforms and accompaniment waveforms respectively, we can redefine \emph{text-to-song} task as the approximation of joint conditional probability optimization $p(S_v, S_a|C, P)$.

\subsection{Overview}

In this work, we propose Melodist, the first controllable text-to-song model. As illustrated in Figure \ref{fig:overview}, it is organized in two stages: 1) In the first stage we follow the common SVS process that generates a singing voice conditioned on the music score; 2) In the second stage we generate musical accompaniments from singing given natural language prompt. Instead of directly modeling distributions over vocal and accompaniment waveforms, we adopt acoustic tokens as the prediction targets. Finally, we reconstruct waveforms from predicted vocal acoustic tokens and accompaniment acoustic tokens and then mix them as the output. 

The fundamental ideas behind the two-stage generation can be summarized as follows: 1) The accompaniment and voice signals inside the same song strongly relate to each other. The vocals and accompaniments are aligned in melody pattern, temporal dynamics, and emotional variation. 2) It reflects the conditional independence assumption that the attribute control applied on accompaniment is independent of the vocals and music score; 3) It is consistent with the dependency that the semantic and acoustic features of the singing voice depend on the music score while the harmony and controllability of accompaniment are decided on vocals and prompts, respectively.

\subsection{Predicted Target}

Acoustic tokens, as the predicted target, are extracted the acoustic tokens by SoundStream \cite{zeghidour2021soundstream}, a neural codec with an encoder-decoder architecture and a residual vector quantizer (RVQ) cascaded $n_q$ layers of vector quantizer (VQ). Assuming $y$ denotes an audio sample, the extracted acoustic tokens sequence can be represented as $Z^{T×n_q} = encoder(y)$ where T refers to the number of frames. These compressed representations can be used to reconstruct waveforms by the decoder subsequently that $\hat{y} = decoder(Z)$.

\subsection{Backbone Model}
We adopt the multi-scale transformer proposed in \cite{yu2023megabyte, yang2023uniaudio} as our backbone in both two stages for long sequence modeling. It also presents outstanding performance in terms of generation and in-context learning capabilities. Specifically, It introduces a hierarchical design consisting of a global transformer and a local transformer, both of which are decoder-only transformers. Specifically, the flattened acoustic token sequence is first chunk into patches $\{x_0, x_1, \dots, x_T\}$ of T frames, each containing $n_q$ tokens of one frame. Let $H$ denote the patch representations, the chunked sequence is passed to the global transformer $G$ to predict the target in a frame-by-frame manner:
\begin{equation}
\begin{aligned}
H^{g\_out}_{1:T} &= G(H^{g\_in}_{0:T-1}), 
\end{aligned}
\end{equation}

In contrast, the local model $L$ operates on a single patch of size $n_q$, each of which is the sum of the output of the global model and the embedding of the previous tokens.
\begin{equation}
\begin{aligned}
H^{l\_out}_{t, 1:n_q} &= L(WH^{g\_out}_{t-1, 0:n_q-1}+H^{l\_in}_{t, 0:n_q-1})
\end{aligned}
\end{equation}

Where W denotes the projection matrix to map the hidden size of the local transformer.

During training, the model is optimized using token prediction and cross-entropy loss. In the inference stage, the model autonomously predicts acoustic tokens in an auto-regressive manner conditioning on prefixed input sequences. Such a design facilitates the reduction of computational and enhances in-context learning for long sequences to a large extent. 

\subsection{Two-stage Synthesis}
\paragraph{Stage 1: Singing Voice Synthesis.} In the SVS stage, the model synthesizes acoustic tokens conditioned on lyric phonemes, durations, and pitch. Specifically, we transform the condition input into discrete tokens and repeat each for $n_q$ times to fill each patch. The expanded inputs and target acoustic tokens are concatenated and embedded into a unified sequence, subsequently processed by the multi-scale transformer.

\paragraph{Stage 2: Vocal-to-accompaniment Synthesis.} In the vocal-to-accompaniment synthesis stage, the model synthesizes acoustic tokens of accompaniment conditioned on vocal acoustic tokens and natural language prompts. We leverage a pre-trained text encoder providing text representation with consistent global characteristics with the vocal and accompaniment, which we will illustrate in section \ref{sec:clap} in detail. It can be incorporated with our backbone model to enhance attribute controllability. We freeze the parameters of the text encoder, utilize it to extract the non-pooled text representation of the prompt, and pass it through a linear layer to fit the dimension of the backbone model. Once we have obtained "continuous text embeddings",  we also repeated each token for $n_q$ times. The inputs are organized and processed in the same way as in the previous stage. 

\begin{figure*}[t]
\centering
\vspace{-3mm}
\includegraphics[width=0.7\textwidth]{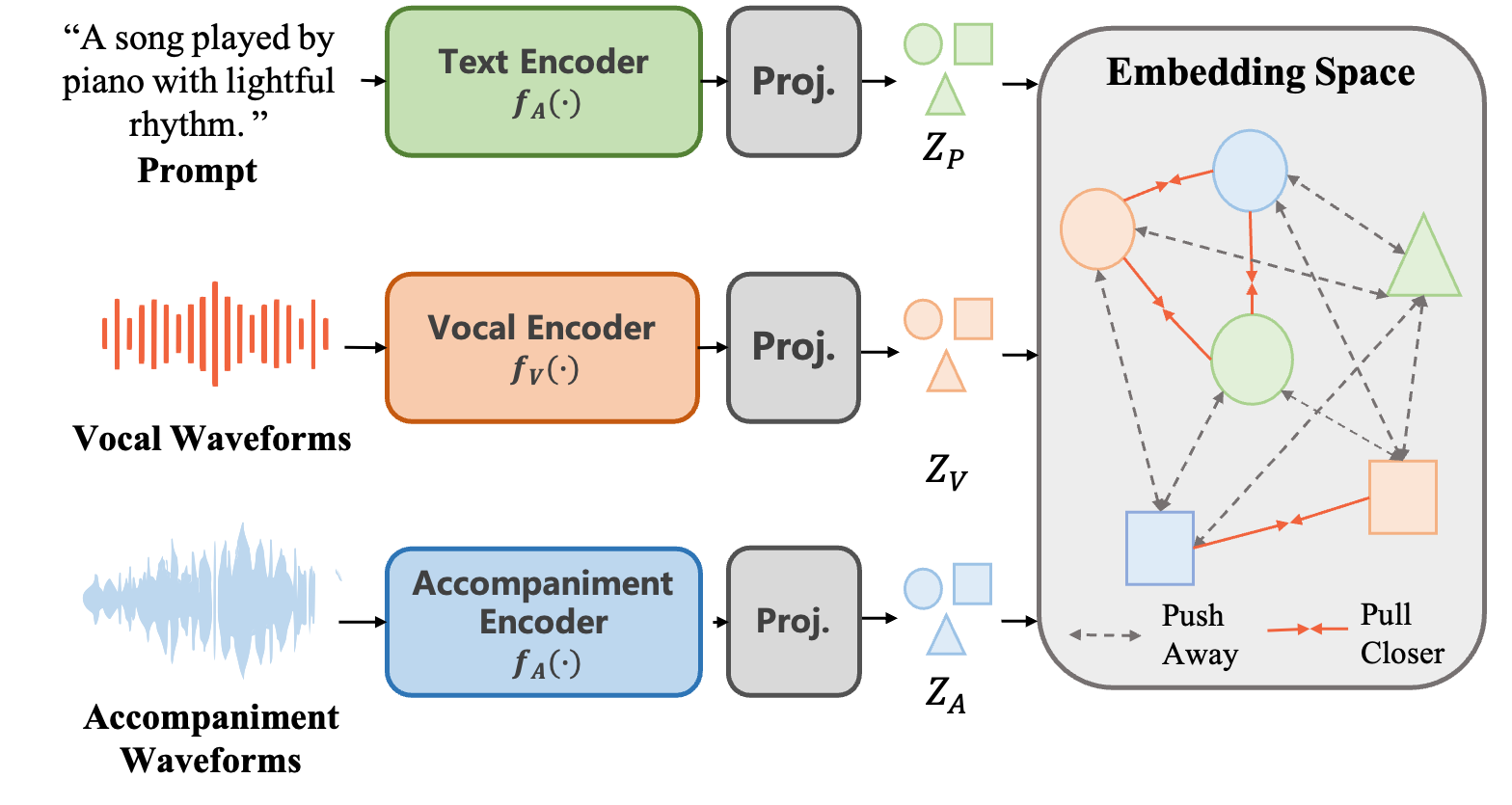}
\caption{The architecture of the tri-tower contrastive framework. $Z_P$, $Z_V$, $Z_A$ refer to the representation extracted by the text encoder, the vocal encoder and the accompaniment encoder, respectively. We use different shapes to represent different triples, while color is used to distinguish the kinds of inputs. Embeddings of the same triplet are pulled closer, while those of different objects are pushed away in the joint embedding space.}
\label{fig:tri}
\end{figure*}
\subsection{Waveform Reconstruction}
Instead of the decoder of Soundstream, we adopt a unit-based vocoder utilizing GAN-based architecture for waveform generation from acoustic units. It is derived from BigvGAN and comprises a generator and two discriminators. Specifically, the generator is built from a set of look-up tables (LUT) that embed the discrete units. It is followed by a series of blocks composed of transposed convolution for the purpose of upsampling and a residual block with dilated layers to expand the receptive field. The multi-period discriminator (MPD) and the multi-resolution discriminator (MRD) proposed in BigvGAN are added to distinguish between the generated audio and ground truth. Note that we train two neural codecs (the vocoders and encoders used to extract acoustic tokens) sharing the same architecture but not the same parameters respectively for vocals and accompaniments. We found that gradient collapse occurs when training only one neural codec on all audios. It is mainly attributed to the distribution discrepancy between the vocals and accompaniments. Once we obtain the waveforms of the vocal and its accompaniments, we mix them in the waveform domain to get the final output.

\section{Tri-Tower Contrastive Pre-training}
\label{sec:clap}
We introduce a tri-tower training scheme with contrastive loss that jointly embeds text, vocals, and accompaniments into an aligned space. As presented in Figure \ref{fig:tri}, it consists of three separate encoders: text encoder $f_P(\cdot)$, vocal encoder $f_V(\cdot)$, and accompaniment encoder$f_A(\cdot)$, each followed by a pooling and linear layer. Parallel text prompt, vocal, and accompaniment make up each triplet of a mini-batch $(x_p, x_v, x_a)$ and they are passed through the respective encoder. The text encoder $f_P(\cdot)$: $\mathcal{A}^n$ $\xrightarrow{}$ $\mathbb{R}^{d_P}$ converts a tokenized text sequence of length n over vocabulary A to the text embedding of dimension d. The Vocal encoder and the accompaniment encoder $f_V(\cdot)$, $f_A(\cdot)$: $\mathbb{R}^{F\times T}$ $\xrightarrow{}$ $\mathbb{R}^{d_V}$ encode log mel spectrograms of the vocal and accompaniment respectively, which F refers to the number of mel channels and T refers to the number of frames. A linear layer is appended in each branch to project the representations into a $l_2$-normalized embedding space.

When considering two-tower contrastive learning, two encoders of different modalities are jointly trained to maximize the similarity between N positive pairs while minimizing the similarity for N $\times (N-1)$ negative pairs. We adopt the multi-modal version of InfoNCE loss \cite{oord2018representation}. Taking pair (text, vocal) as an example, the loss can be formulated as follows:
\begin{equation}
\begin{aligned}
L_{p\rightarrow{}v}&=-\log \frac{\exp(z_{p_i}\cdot z_{v_i}/\tau)}{\sum_{j=1}^N \exp(z_{p_i}\cdot z_{v_j}/\tau)}
\end{aligned}
\end{equation}
\begin{equation}
\begin{aligned}
L_{p\leftrightarrow{}v} &= (L_{p\rightarrow{}v} + L_{v\rightarrow{}p})/2
\end{aligned}
\end{equation}
Where $\tau$ is a temperature parameter. To extend it into tri-tower contrastive loss, we simply calculate the contrastive loss over pairs of the representations in a triplet (text, vocal, accompaniment) that:
\begin{equation}
\begin{aligned}
L &= L_{p\leftrightarrow{}v}+L_{p\leftrightarrow{}a}+L_{v\leftrightarrow{}a}
\end{aligned}
\end{equation}

To verify the effectiveness of the tri-tower contrastive pre-training framework, we also compare it with CLAP on two related cross-modal retrieval tasks: text-vocal retrieval and text-accompaniment retrieval. We report the experimental results in section \ref{sec:retrieval}, which indicates that including both vocal and accompaniment helps the model learn to ground more attribute-related song concepts.  
\section{Experiments}
\subsection{Dataset}
To our knowledge, there are no public datasets available for controllable text-to-song. We crawl five thousand Mandarin songs covering around fifty singers, their lyrics, and some attribute tags (mood, instruments, style, etc.) from a well-known music website. There are 180 hours of audio data in total. In order to get the desired input, we perform some filtering and processing operations on the data. We present the details of data analysis and processing in Appendix \ref{sec:crawldataset}.

To alleviate data scarcity, we also leverage some open-source Mandarin singing voice datasets, which are listed in Appendix \ref{sec:dataset}.

\subsection{Training and Evaluation}
\paragraph{Model Configurations.}For the tri-tower contrastive learning framework, we adopt the base version of BERT \cite{devlin2018bert} as the text encoder and the modified version of Audio Spectrogram Transformer \cite{gong2021ast} as the architecture of vocal encoder and accompaniment encoder. The [CLS] token from the final layer is projected into the joint embedding space of size 128. SoundStream \cite{zeghidour2021soundstream} has 12 quantization levels, each with a codebook of 1024 entries. The first three quantization levels are employed as acoustic tokens. The generator of unit-based vocoder is built from the modified V1 version of BigVGAN \cite{bigvgan}. A comprehensive illustration of model hyperparameters is available in Appendix \ref{model}.

\paragraph{Experimental Setup.}We apply Spectrogram augmentation and text augmentation strategies for better performance. It takes 30 epochs for tri-tower pre-training using 8 NVIDIA V100 GPUs with a batch size of 128. For the training of text-to-song synthesis, we train the SVS model for 80K steps and the vocal-to-accompaniment model for 60K steps, both using 6 NVIDIA V100 GPUs with a batch size of 5000 tokens for each GPU. Each unit-based vocoder is trained using 4 NVIDIA V100 GPUs for 150K steps until convergence. The detailed setup is presented in Appendix \ref{exp}.

\begin{table}[t]
\centering
\resizebox{\linewidth}{!}{
\begin{tabular}{lccc}
\toprule
\textbf{Model}       & \textbf{MOS ($\uparrow$)} & \textbf{SMOS ($\uparrow$)} & \textbf{FFE ($\downarrow$})  \\ \midrule
GT           & 4.02$\pm$ 0.05   & /  & /   \\ \midrule
FFT-Singer   & 3.71$\pm$ 0.08   & 3.79$\pm$0.07  & 0.20 \\
DiffSinger   & 3.80$\pm$ 0.06   & 3.85$\pm$0.08 & 0.18 \\
VISinger   & 3.82$\pm$ 0.05   & 3.86$\pm$0.05 & 0.15 \\
Make-A-Voice   & 3.86$\pm$ 0.04  &\textbf{3.89$\pm$0.08} & 0.11 \\
\textbf{Melodist} & \textbf{3.90$\pm$0.06} &3.87$\pm$0.07 &\textbf{0.09}  \\
\bottomrule
\end{tabular}}
\caption{Ojective and subjective evaluation for Melodist and SVS baselines.}
\label{tab:svs}
\end{table}
\paragraph{Evaluation.}We conduct both subjective and objective evaluations on generated samples. 

Regarding the evaluation of SVS synthesis, we conduct a crowd-sourced human evaluation via Amazon Mechanical Turk on the metrics of Mean Opinion Score (MOS) and Similarity Mean Opinion Score (SMOS) both with 95 \% confidence intervals, which measures sample quality and speaker similarity respectively. We also calculate the F0 Frame Error (FFE) for objective evaluation. 

Regarding the evaluation of accompaniment synthesis, we asked the raters to evaluate the audio samples in terms of overall quality (OVL), relevance to the prompt (REL), and alignment with the melody (MEL.). For the objective evaluation, we calculate the Fr$\acute{e}$chet Audio Distance (FAD), Kullback–Leibler Divergence (KLD), and the CLAP score (CLAP). We have attached the setting of evaluation in Appendix \ref{sec:eval}.

\subsection{Singing Voice Synthesis}
We compare our SVS model with four recent SVS baselines: 1) FFT-Singer, which generates mel-spectrograms through stacked feed-forward transformer blocks; 2) DiffSinger \cite{diffsinger}, which was built on diffusion probabilistic models to generate mel-spectrograms; 3) VISinger \cite{visinger}, an end-to-end singing synthesis model 4) Make-A-Voice \cite{make-a-voice}, a multimodal spoken large language model for synthesizing and manipulating voice signals. We also train a BigvGAN vocoder on 16k audios for FFT-Singer and DiffSinger to reconstruct waveform from Mel-spectrograms. 

As shown in Table \ref{tab:svs}, our SVS model outperforms other baseline models with the highest MOS score of 3.89, indicating that it enjoys great superiority in sample quality. The SMOS score lags behind that of Make-A-Voice by a narrow margin but is better than other baseline models. The highest FFE score demonstrates the proficiency of Melodist in emulating the pitch prompt.

\begin{table*}[ht]
\centering
\tiny
\vspace{-6mm}
\resizebox{\textwidth}{!}{
\begin{tabular}{lccccccc}
\toprule
\textbf{Model}       & \textbf{FAD ($\downarrow$)} & \textbf{KLD ($\downarrow$)} & \textbf{CLAP ($\uparrow$)}  & \textbf{OVL. ($\uparrow$)} & \textbf{REL. ($\uparrow$)} &\textbf{MEL ($\uparrow$)}\\ \midrule
MUSICGEN (T5) & 4.28   & 1.48 & 0.27 & 81.12$\pm$1.34 &83.06$\pm$1.70 &67.72$\pm$1.23 \\
MUSICGEN (CLAP)   & 4.97   & 1.61 & 0.33 & 78.64$\pm$1.02 &85.01$\pm$1.43 &61.29$\pm$0.83 \\ \midrule
Melodist (T5)   & \textbf{3.69}   & 1.36  & 0.29 &\textbf{83.87$\pm$1.23} &83.58$\pm$1.61 &78.05$\pm$0.75 \\ 
Melodist (CLAP)   & 4.10   & 1.59 & 0.34 &78.75$\pm$1.54 &85.19$\pm$1.23 &70.33$\pm$0.92 \\                                                 
Melodist (Tri-Tower)  & 3.80  & \textbf{1.34} & \textbf{0.39}  &83.15$\pm$1.46 &\textbf{86.63$\pm$1.27} &\textbf{79.40$\pm$0.96} \\ 
\bottomrule
\end{tabular}}
\caption{Objective and Subjective evaluation of accompaniment samples generated by Melodist and MUSICGEN.}
\label{tab:axy}                 
\end{table*}
\subsection{Vocal-to-accompaniment Synthesis}

\subsubsection{Comparison to baselines}
To our knowledge, SingSong \cite{donahue2023singsong} is the only model with the same experimental setup as ours. However, its code and dataset are not available. So we only compare our model with MUSICGEN \cite{copet2023musicgen}, a controllable music generation model that can be conditioned on text and melody. Specifically, we adopt the vocal track extracted by Demucs as the melody condition of MUSICGEN. As reported in Table \ref{tab:axy}, we also investigate the impact of different text encoder including: 1) T5 \cite{raffel2020t5}, which is a Transformer architecture using a text-to-text approach; and 2) CLAP \cite{elizalde2023clap}, a model for learning audio concepts from natural language supervision. 

In general, Melodist surpasses MUSICGEN in objective and subjective metrics when applying the same text encoder, indicating the superiority of flattening prediction compared to the codebook interleaving strategies proposed in MUSICGEN. It reaches a trade-off between performance and computational efficiency. 

Melodist presents the highest perceptual quality with outperformed FAD and OVL evaluation. When equipped with the text encoder of the tri-tower framework, the FAD and OVL scores drop slightly but still present better performance compared to MUSICGEN. 

The adherence to the prefix condition can be witnessed in the evaluation result. Regarding to text prompts, Melodist outperforms MUSICGEN with the highest CLAP and REL scores and the lowest KLD score. Regarding to melody evaluation, the experimental results suggest that Melodist scores the best alignment with the melody of input, indicating that it can successfully generate accompaniments in harmony with the singing voice in melody. 

\subsubsection{Comparison of Different Text Encoder}
 
The evaluation results are reported in Table \ref{tab:axy}. In terms of adherence to text prompts, the Tri-tower framework outperforms other text encoders with the highest CLAP and REL score and the lowest KLD score. The superiority of the Tri-tower framework can be witnessed. It indicates that Melodist is capable of generating accompaniments that share similar semantic concepts with the text prompts while ensuring favorable audio quality. It can be observed that the text encoders trained in the contrastive learning paradigm show a better alignment between generated audios and text prompts, which demonstrates that the contrastive pre-training scheme significantly enhances text-guided music generation. However, there is a subtle gap in terms of audio quality, as reflected in the slightly worse FAD and OVL score. The discrepancy can be mainly attributed to model capacity and the pre-training objective.

\begin{table*}[t]
\centering
\tiny
\vspace{-3mm}
\resizebox{\textwidth}{!}{
\begin{tabular}{lccccccc}
\toprule
\textbf{V2A Model}     
& \textbf{FAD ($\downarrow$)} & \textbf{KLD ($\downarrow$)} & \textbf{CLAP ($\uparrow$)}  & \textbf{OVL. ($\uparrow$)} & \textbf{REL. ($\uparrow$)} &\textbf{MEL ($\uparrow$)}\\ \midrule
MUSICGEN  & 3.97   & 1.39 & 0.27 & 82.33$\pm$1.05 &82.92$\pm$1.45 &65.08$\pm$0.74 \\ 
Melodist  & \textbf{3.81}  & \textbf{1.34} & \textbf{0.39}  &\textbf{84.28$\pm$1.70} &\textbf{85.72$\pm$1.29} &\textbf{75.86$\pm$1.06} \\ 
\bottomrule
\end{tabular}}
\caption{Objective and Subjective evaluation of song samples generated by Melodist and MUSICGEN.}
\label{tab:axy_song}                 
\end{table*}

\subsubsection{Cross-modal Retrieval Result}
\label{sec:retrieval}

To further verify the effectiveness of the tri-tower contrastive framework, We conduct experiments of text-vocal retrieval and text-accompaniment retrieval. Specifically, we use 1K recordings as the pool of candidates and the paired vocal or accompaniment as the ground truth. We compare our tri-tower contrastive framework with three baselines: 1) MusCALL \cite{manco2022contrastive}, a contrastive audio-language framework for Music; 2) MULAN \cite{huang2022mulan}, a music audio and natural language joint embedding model; 3) CLAP \cite{elizalde2023clap}, a model for learning audio concepts from natural language supervision. The sentence-level retrieval performance is evaluated by: 1) measuring mean average precision (mAP) for accuracy evaluation; and 2) Recall at the top k ranks (Recall@k). We set k to 1, 5, and 10. 

\begin{table}[!b]
\tiny
\centering
\resizebox{\linewidth}{!}{
\begin{tabular}{lcccc}
\toprule
\multirow{2}{*}{\textbf{Model}}        & \multicolumn{3}{c}{\textbf{Recall ($\uparrow$)}}& \multirow{2}{*}{\textbf{mAP ($\uparrow$)}}\\ \cmidrule{2-4}
& \textbf{@1} & \textbf{@5} & \textbf{@10} &   \\ \midrule
\multicolumn{5}{l}{\textbf{Text-to-vocal Retrieval}}\\ \midrule
MusCALL & 6.5   & 20.6  & 31.3  & 12.2 \\ 
MULAN   & 8.2   & 22.7  & 34.5 & 13.0 \\
CLAP  & 5.4   & 17.9  & 29.6 & 9.8 \\
\textbf{Melodist} & \textbf{9.8} & \textbf{25.1} & \textbf{40.4}  & \textbf{16.3} \\ \midrule
\multicolumn{5}{l}{\textbf{Text-to-accompaniment Retrieval}} \\ \midrule
MusCALL  & 7.4   & 23.1  & 36.0  & 13.9 \\ 
MULAN   & 8.0   & 22.3  & 38.2 & 15.3 \\
CLAP   & 6.8   & 21.5 & 36.9 & 13.0 \\
\textbf{Melodist} & \textbf{11.2} & \textbf{28.0} & \textbf{43.9}  & \textbf{19.4}\\
\bottomrule
\end{tabular}}
\caption{The experimental results of text-vocal retrieval and text-accompaniment retrieval.}
\label{tab:retrieval}
\end{table}

As presented in Table \ref{tab:retrieval}, a significant superiority can be observed from these recall rates and the mean average precision, indicating that including both vocal and accompaniment helps the model learn to ground more attribute-related song concepts. Jointly learning from vocals and accompaniments facilitates the text encoder extracting more accurate text representations of global characteristics, which greatly assists in subsequent vocal-to-accompaniment modeling. In addition, it is interesting that better retrieval performance is presented in text-to-accompaniment retrieval. This is mainly due to the reason that the text descriptions are more relevant to the accompaniment.

\subsection{Text-to-song Synthesis}
After a stage-by-stage evaluation, we compare the songs generated by Melodist and MUSICGEN in general terms. We fix the singing voice synthesis stage and generate the accompaniments with MUSICGEN and Melodist respectively. The only difference lies in the vocal-to-accompaniment model used for vocal-to-accompaniment synthesis. As we can see in Table \ref{tab:axy_song}, Melodist presents the highest perceptual quality and the best adherence to text prompt. It is identical to the observation of the previous section that Melodist outperforms MUSICGEN with outperformed scores, which is identical to the observation of the previous section.

\subsection{Ablation}
In this section, we investigate the impact of different data combinations and different augmentation strategies. Details and experimental results of the ablation can be found in the Appendix \ref{sec:abl}.
\paragraph{Data Combination.} We consider four combinations of crawled data and open-source data. We found that the absence of open-source SVS data leads to worse SVS performance, while a noticeable performance degradation in terms of audio quality and adherence can be witnessed when excluding open-resource song data. 
\paragraph{Data Augmentation Strategies.} We explore the effectiveness of text augmentation and spectrogram augmentation. When analyzing the experimental results, we can see a decline in both recall and mAP scores. A noticeable gain can be witnessed when applying data augmentation strategies.

\section{Conclusion}
In this paper, we introduce a new task called \emph{text-to-song}, which incorporates singing voice and accompaniment synthesis from music score. We propose Melodist, the first text-to-song model with a two-stage generation scheme. Natural language prompts serve as the condition to control accompaniment generation. Melodist leverage a tri-tower contrastive pre-training framework to align the prompt with its vocal and accompaniment. We have collected a Mandarin song dataset from the music website and leverage some open-source song and singing datasets to alleviate the data scarcity. We have conducted a series of comprehensive evaluations and the results indicate that Melodist outperforms baselines with comparable audio quality, temporal correspondence, and consistency with text concept. We provide extensive experiments to demonstrate the effectiveness of the tri-tower contrastive learning framework as well as the impact of different data combination and data augment strategies. In the future, we will focus on improving the audio quality and vocal accompaniment harmonization.
\clearpage
\section*{Limitations}
Though Melodist have shown comparable achievements, its limitations cannot be ignored. The reliance on source separation imposes a great challenge to improving audio quality. While the current source separation methods remain suboptimal, it is urgent to improve the quality of source separation. There are some alternatives such as constructing a high-quality dataset or designing a fully end-to-end text-to-song synthesis model. Additionally, Melodist treats accompaniment as a single track, disregarding the intricate composition of individual elements such as drums, bass, and other instrument-related tracks. A promising avenue for future exploration involves both intra-track and inter-track modeling, thereby facilitating a more comprehensive approach to text-to-song synthesis.


\bibliography{reference}

\clearpage
\appendix
\section{The Details of Experiment}
\subsection{Model Configuration}
\label{model}
The model hyper-parameters of Melodist are listed in Table \ref{tab:melodist}.

\begin{table*}[!h]
\centering
\tiny
\resizebox{0.8\linewidth}{!}{
\begin{tabular}{l|c|c|c}
\toprule
\multicolumn{2}{c|}{\textbf{Hyperparameter}}   & \textbf{Melodist} & \textbf{Number of parameters} \\ 
\midrule
\multirow{4}{*}
{\makecell[c]{Global\\Transformer}}
& Hidden Size    &192 & \multirow{4}{*}{320.07M}\\
& Layers          &20      \\
& Hidden Dim      &1152     \\                  
& Attention Heads     &16   \\    
& FFN Dim             &4608\\ 
\midrule
\multirow{4}{*}
{\makecell[c]{Local\\Transformer}}
& Hidden Size  &192 & \multirow{4}{*}{100.14M}  \\
& Layers          &6      \\
& Hidden Dim      &1152     \\                  
& Attention Heads     &8   \\    
& FFN Dim             &4608\\
\midrule
\multirow{3}{*}
{\makecell[c]{Unit-based\\Vocoder}}
& Upsample Rates  &[5, 4, 2, 2, 2, 2] & \multirow{3}{*}{121.60M}  \\
& Hop Size         &320      \\
& Upsample Kernel Sizes      &[9, 8, 4, 4, 4, 4]  \\
\midrule
\multirow{4}{*}
{\makecell[c]{Vocal Encoder}}
& Layers   &6 & \multirow{4}{*}{42.10M}  \\
& Hidden Dim         &768      \\
& Attention Heads      &8  \\
& FFN Dim             &3072\\
\midrule
\end{tabular}}
\caption{Hyperparameters of Melodist.}
\label{tab:melodist}
\end{table*}

\subsection{Experimental Setup}
\label{exp}
In Tri-tower contrastive pretraining, each audio is converted to a log-scaled mel spectrogram with the FFT size of 1024, hop size of 256, and window size of 1024. We then chunk the augmented spectrogram into 16$\times$ 16 patches. We limit the max text sequence length to 77 chars for computational efficiency. Inspired by \cite{copet2023musicgen}, text augmentation is applied by concatenating tag lists to the text description. We limit the max text sequence length to 77 chars for computational efficiency. A [CLS] token is prepended to the sequence as a summary of the contextual patch embeddings in three encoders. We set the temperature $\tau$ to 0.2. 

For two-stage text-to-song synthesis, the learning rate is set to 5e-5. Adam optimizer is used with $\beta_1=0.9$, $\beta_2=0.98$, and $\epsilon=10^{-9}$. We use Top-k sampling for inference, in which k and the temperature are set to 30 and 0.8. 

The unit-based vocoder is trained on 16k audio data with a segment size of 32000. The learning rate is set to 5e-5. Adam optimizer is used with $\beta_1=0.8$, $\beta_2=0.99$, and $\epsilon=10^{-6}$.

\section{Dataset Analysis}
In this section, we describe the details of the dataset for training. 
\subsection{Open-Source Datasets}
\label{sec:dataset}
We present the open-source datasets adopted for training in Table \ref{tab:dataset}.
\begin{table*}[t]
\centering
\resizebox{0.8\linewidth}{!}{
\begin{tabular}{lccc}
\toprule
\textbf{Dataset}    &\textbf{Type}  &\textbf{Annotation}  &\textbf{Volume (hrs)}  \\ \midrule
\multicolumn{4}{l}{\textbf{Stage 1: Singing Voice Synthesis}} \\ \midrule 

Opencpop \cite{wang2022opencpop}   &singing   &text, duration, MIDI  & 5.2 \\
M4Singer \cite{zhang2022m4singer}  &singing   &text, duration, MIDI & 29.8 \\
OpenSinger \cite{opensinger}  &singing   &text, duration, MIDI  & 86.5 \\
PopCS \cite{diffsinger}  &singing   &text, duration   & 5.9 \\
AISHEELL-3 \cite{aishell3}             &speech   &text   & 85 \\ \midrule
\multicolumn{4}{l}{\textbf{Stage 2: Vocal-to-accompaniment Synthesis}} \\ \midrule 
LP-MusicCaps-MSD \cite{lp-musiccaps-msd}  &music   &text description   & 7k \\
\bottomrule
\end{tabular}}
\caption{Statistics of training datasets.}
\label{tab:dataset}
\end{table*}

\subsection{The Crawled Song Data}
\label{sec:crawldataset}
\subsubsection{Data processing pipeline}
In order to get the desired input, we perform the following filtering and processing operations on the data:
\paragraph{Data Filtering.} We exclude audios that 1) are live songs; 2) of silent accompaniment or no vocals; 3) are performed by multiple singers. Additionally, some content (composer, performer, etc.) irrelevant to text transcriptions is removed from the lyrics.

\paragraph{Source Separation.} We split each song into 10-second clips from each song and passed each clip to the Demucs \cite{demucs} to separate vocals from the rest of the accompaniments and yield aligned pairs of waveforms. Finally, we resample vocal and instrumental clips from 44.1kHz to 16kHz and average all audio files to mono.

\paragraph{Lyrics-to-Singing Alignment.} We first reorganize the clips of the separated vocals and restore them to the original songs. Then we use Montreal forced alignment \cite{mcauliffe2017montreal} tool to extract the phoneme duration. After filtering the misaligned segments, we segment each song in 6-10s according to the separation marks in raw lyrics. 

\paragraph{Pitch Extraction.} We extract F0 (fundamental frequency) from the raw waveform using Parselmouth to provide pitch information. We have quantified F0 to its rounded value.

\paragraph{Prompt Generation.} We copy the tags of a song to its segments and then make minor modifications according to the auditory impression. A tag-to-pseudo caption generation approach with large language models \cite{lp-musiccaps-msd} is leveraged to generate natural language prompts.

\subsubsection{Examples of Prompt}
We provide some examples of attribute tag lists and the captions generated by \cite{lp-musiccaps-msd}.

There are examples of crawled attribute tag lists:
\begin{itemize}
[left=0em,label=\textbullet]
    \setlength{\itemindent}{0pt}
    \setlength{\tabcolsep}{0pt}
    \setlength{\parskip}{0pt}
    \setlength{\partopsep}{0pt}
    \setlength{\itemsep}{0pt}
    \setlength{\topsep}{0pt}
    \setlength{\parsep}{0pt}
    \item pop, bass, guitar, acoustic, beat.
    \item rock, passionate, vocal, shimmering, bass, guitar, acoustic, guitar, guitar, emotional, passionate.
    \item instrumental, melodic, saxophone, acoustic, guitar, soft, mellow, ambient, dreamy.
    \item cool, vocal, bass, percussion, retro, dance.
    \item guitar, synth, bass, guitar, electronic, beat, sentimental, dance, club
\end{itemize}

There are examples of generated text descriptions:
\begin{itemize}
[left=0em,label=\textbullet]
    \setlength{\itemindent}{0pt}
    \setlength{\tabcolsep}{0pt}
    \setlength{\parskip}{0pt}
    \setlength{\partopsep}{0pt}
    \setlength{\itemsep}{0pt}
    \setlength{\topsep}{0pt}
    \setlength{\parsep}{0pt}
    \item This is a pop music piece. There is a male vocalist singing melodically in the lead. The melody is being played by the keyboard while the bass guitar is playing in the background. The rhythm consists of a slow tempo electronic drum beat. The atmosphere is easygoing. This piece could be used in the soundtrack of a romantic comedy movie, especially during the scenes where a character is hesitating to open up to their crush.
    \item The low quality recording features a rock song that consists of a passionatele vocal singing over punchy kick and snare hits, shimmering hi hats, soft kick and groovy bass guitar. It sounds addictive, energetic and passionate.
    \item This music is a Jazz instrumental. The tempo is slow with a melodic saxophone harmony, keyboard accompaniment and rhythmic acoustic guitar accompaniment. The music is soft, mellow, pleasant, ambient, dreamy and pleasant.
    \item A female singer sings this cool melody with backup singers in vocal harmony. The song is medium tempo with a steady drumming rhythm, keyboard accompaniment, percussive bass line and various percussion hits. The track is a retro hip hop dance tune.
    \item This is an amateur recording of a R\&B music piece. There is a male vocalist singing melodically in the lead. The melody is being played by the electric guitar and the synth bass guitar while the rhythmic background consists of a slow tempo electronic drum beat. The atmosphere is sentimental. This piece could be playing in the background at a nightclub or a dance club.
\end{itemize}

\begin{table*}[!ht]
\centering
\resizebox{\linewidth}{!}{
\begin{tabular}{c|c|c|cc|cccc}
\hline
\multirow{2}{*}{\textbf{ID}} & \multirow{2}{*}{\textbf{SVS Data}} & \multirow{2}{*}{\textbf{Song Data}} &\multicolumn{2}{c|}{\textbf{Stage 1}} &
\multicolumn{4}{c}{\textbf{Stage 2}} \\ \cline{4-9}
& & & \textbf{MOS} ($\uparrow$) & \textbf{FFE} ($\downarrow$) & \textbf{FAD ($\downarrow$)} & \textbf{KLD} ($\downarrow$) & \textbf{OVL ($\uparrow$)} & \textbf{REL ($\uparrow$)}\\ \hline
1  &\ding{51} &\ding{55}   & 3.89$\pm$0.08  & 0.09 & 3.88  & 1.46 &79.56$\pm$1.42 &83.02$\pm$1.39 \\
2  &\ding{55} &\ding{51}   & 3.84$\pm$0.05   & 0.13 & 3.79 &1.39 &83.10$\pm$1.31 &86.56$\pm$1.80 \\
3  &\ding{55} &\ding{55}   & 3.84$\pm$0.05 & 0.13 & 3.88  & 1.46 &79.56$\pm$1.42 &83.02$\pm$1.39\\ 
4  &\ding{51} &\ding{51}   & \textbf{3.89$\pm$0.08} & \textbf{0.09} &\textbf{3.79} &\textbf{1.39} &\textbf{83.10$\pm$1.31} &\textbf{86.56$\pm$1.80} \\ 
\midrule                     
\end{tabular}}
\caption{Ablation study on different data combination.}
\label{tab:abl_data}
\end{table*}

\section{Evaluation}
\label{sec:eval}

\section{Ablation Study}
\label{sec:abl}

\begin{table}[!b]
\tiny
\centering
\resizebox{\linewidth}{!}{
\begin{tabular}{lcccc}
\toprule
\multirow{2}{*}{\textbf{Model}}        & \multicolumn{3}{c}{\textbf{Recall ($\uparrow$)}}& \multirow{2}{*}{\textbf{mAP ($\uparrow$)}}\\ \cmidrule{2-4}
& \textbf{@1} & \textbf{@5} & \textbf{@10} &   \\ \midrule
\multicolumn{5}{l}{\textbf{Text-to-vocal Retrieval}}\\ \midrule
w/o TA & 6.7   & 18.2  & 34.2  & 13.7 \\ 
w/o SA   & 8.0   & 20.6  & 33.9 & 12.2 \\
w/o TA\&SA    & 6.3   & 15.8  & 32.3 & 10.3 \\
\textbf{TA\&SA} & \textbf{9.8} & \textbf{23.7} & \textbf{40.2}  & \textbf{15.7} \\ \midrule
\multicolumn{5}{l}{\textbf{Text-to-accompaniment Retrieval}} \\ \midrule
w/o TA  & 7.4   & 21.1  & 37.0  & 14.5 \\ 
w/o SA   & 8.5   & 22.3  & 39.1 & 15.9 \\
w/o TA\&SA    & 6.2   & 18.5  & 35.9 & 13.1 \\
\textbf{TA\&SA} & \textbf{11.3} & \textbf{27.6} & \textbf{41.1}  & \textbf{19.4}\\
\bottomrule
\end{tabular}}
\caption{Ablation study on the impact of data augmentation strategies. We report the experimental results of text-vocal retrieval and text-accompaniment retrieval. SA denotes spectrogram augmentation and TA denotes text augmentation.}
\label{tab:abl_retrieval}
\end{table}

\subsection{Comparison with MUSICGEN}
We report objective metrics on the unbalanced set of MusicCaps benchmark, while we sample examples from our crawled dataset. The VGGish, Patchout and CLAP model used for objective evaluation is consistent with \cite{copet2023musicgen}. 

\subsection{Subjective Evaluation}
We randomly selected 30 audio samples generated from each stage and each sample was evaluated by 20 raters via Amazon Mechanical Turk. We paid \$8 an hour for participant compensation.

For quality evaluation of generated singing voice, we conduct the MOS (mean opinion score) tests and explicitly instruct the raters to “(focus on examining the audio quality and naturalness, and ignore the differences of style (timbre, emotion, and prosody).)”. The testers present and rate the samples, and each tester is asked to evaluate the subjective naturalness on a 1-5 Likert scale.

For speaker similarity evaluation, we ask the raters to focus on the similarity of the speaker identity (timbre) to the reference and ignore the differences in content, grammar, or audio quality. We paired each synthesized utterance with a reference utterance to evaluate how well the synthesized speech matched that of the target speaker.

For the evaluation of generated accompaniments, we follow \cite{copet2023musicgen,huanggenerspeech} to evaluate overall quality (OVL), and relevance to the text input (REL). In terms of alignment with the melody (MEL.), we ask the rates to focus more on temporal correspondence between accompaniment and reference singing voice instead of melody resemblance. 

\begin{figure}[!hb]
\centering
\includegraphics[width=\textwidth]{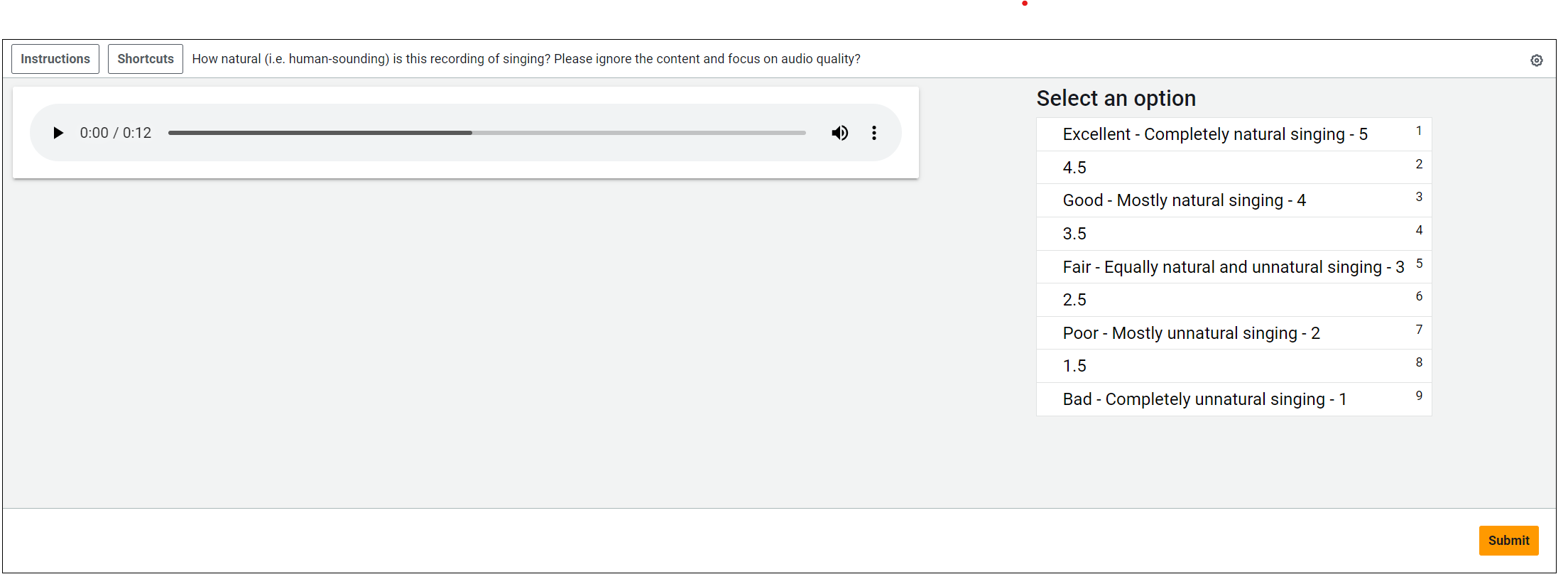}
\caption{Screenshot of MOS testing.}
\label{fig:mos}
\end{figure}

\begin{figure}[!hb]
\centering
\includegraphics[width=\textwidth]{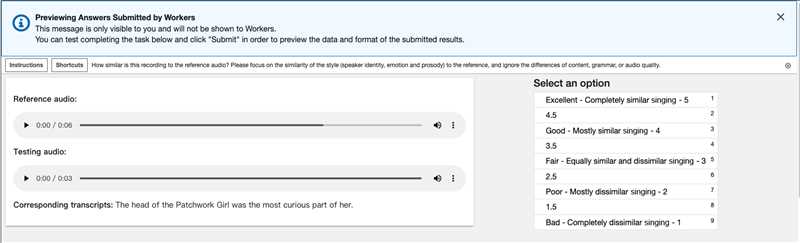}
\caption{Screenshot of SMOS testing.}
\label{fig:smos}
\end{figure}

The Screenshot of subjective evaluation is presented in Figure \ref{fig:mos}, \ref{fig:smos}. A small subset of samples used in the test is available at \url{https://research.github.io/text-to-song/}.

\paragraph{Data Combinations.} We consider four combinations of crawled data and open-source data. when training Melodist, including 1) Exclude open-source SVS data in Stage 1; 2) Exclude song data in Stage 2; 3) Exclude open-source SVS and song data; 4) Include open-source SVS and song data as the original setting.

\paragraph{Data Augmentation.} 
We explore the effectiveness of text augmentation and spectrogram augmentation. 

We report the evaluation results in Table \ref{tab:abl_data} and Table \ref{tab:abl_retrieval}. It suggests that leveraging open-source datasets and augmentation strategies enhance the capability of Melodist to generate more high-fidelity and consistent output.

\clearpage

\end{document}